\documentclass[a4paper, amsfonts, amssymb, amsmath, reprint, showkeys, nofootinbib, twoside]{revtex4-1}
\setcitestyle{super}
\pdfoutput=1
\usepackage[english]{babel}
\usepackage[utf8]{inputenc}
\usepackage[colorinlistoftodos, color=green!40, prependcaption]{todonotes}
\usepackage{amsthm}
\usepackage{mathtools}
\usepackage{physics}
\usepackage{xcolor}
\usepackage{graphicx}
\usepackage[caption=false]{subfig}
\usepackage[left=23mm,right=13mm,top=35mm,columnsep=15pt]{geometry} 
\usepackage{adjustbox}
\usepackage{placeins}
\usepackage[T1]{fontenc}
\usepackage{lipsum}
\usepackage{csquotes}
\usepackage[pdftex, pdftitle={Article}, pdfauthor={Author}, hidelinks]{hyperref} 
\usepackage{physics}
\usepackage{bm}
\usepackage{multirow}
\usepackage{array}
\usepackage{siunitx}
\usepackage{svg}
\newcolumntype{P}[1]{>{\centering\arraybackslash}p{#1}}
\newcolumntype{M}[1]{>{\centering\arraybackslash}m{#1}}
\usepackage[acronym]{glossaries}
\usepackage[inline]{enumitem}
\glsdisablehyper

\newacronym{sidb}{SiDB}{silicon dangling bond}
\newacronym{cad}{CAD}{computer-aided design}
\newacronym{fcn}{FCN}{field-coupled nanocomputing}
\newacronym{qca}{QCA}{quantum-dot cellular automata}
\newacronym{bdl}{BDL}{binary-dot logic}
\newacronym{cmos}{CMOS}{complementary metal-oxide-semiconductor}

\newcommand{\etal}{\textit{et al}.}

\newcommand{\hsi}{H-Si(100)2$\times$1}

\begin{document}

\title{Automated Atomic Silicon Quantum Dot Circuit Design via\\Deep Reinforcement Learning}

\author{Robert~Lupoiu$^{1 \ast}$}
\author{Samuel~S.~H.~Ng$^2$}
\author{Jonathan A. Fan$^1$}
\author{Konrad Walus$^2$}

\address{
1.Department of Electrical Engineering, Stanford University, Stanford, CA, USA \\ 
2.Department of Electrical and Computer Engineering, University of British Columbia, Vancouver, BC, Canada}

\author{$^{\ast}$Corresponding author: \emph{rclupoiu@stanford.edu}}

\maketitle

\section{Abstract}

Robust automated design tools are crucial for the proliferation of any computing technology. We introduce the first automated design tool for the silicon dangling bond quantum dot computing technology, which is an extremely versatile and flexible single-atom computing circuitry framework. The automated designer is capable of navigating the complex, hyperdimensional design spaces of arbitrarily sized design areas and truth tables by employing a tabula rasa double-deep Q-learning reinforcement learning algorithm. Robust policy convergence is demonstrated for a wide range of two-input, one-output logic circuits and a two-input, two-output half-adder, designed with an order of magnitude fewer SiDBs in several orders of magnitude less time than the only other half-adder demonstrated in the literature. We anticipate that reinforcement learning-based automated design tools will accelerate the development of the SiDB quantum dot computing technology, leading to its eventual adoption in specialized computing hardware.

\section{Introduction}



Recent demonstration of nanoscale logic components composed of \glspl{sidb} acting as quantum dots \cite{huff2018binary} signal the emergence of a new contender in the search for successors to \gls{cmos} devices as logic building blocks. These \glspl{sidb} are fabricated on the hydrogen-passivated Silicon(100)-2$\times$1 (\hsi{}) surface with an n-doped bulk and a near surface depletion region \cite{haider2009controlled, huff2018binary, rashidi2018initiating}. Discrete charge states can be observed in \glspl{sidb} \cite{haider2009controlled, pitters2011tunnel}, including positive, neutral, and negative charge states corresponding to the localization of 0, 1, and 2 electrons respectively. An n-doped bulk leads to the tendency for \glspl{sidb} to take on negative charge states in isolation; charges may also be shared between very closely situated \glspl{sidb} \cite{pitters2011tunnel, rashidi2018initiating}. \glspl{sidb} can be created at atomically precise locations by the desorption of individual hydrogen atoms at the surface with electric currents applied by the probes of scanning tunneling microscopes \cite{achal2018lithography, huff2017atomic, pavlicek2017tipinduced}. They can also be individually erased by a mechanically induced passivation of the \gls{sidb} by bringing a functionalized probe very close to the \gls{sidb}\cite{huff2017atomic}. \glspl{sidb} are also known as \textit{atomic silicon quantum dots} in the context that this work is interested in.

The ability to create \glspl{sidb} at atomically precise locations and the exhibited tendency for charge configurations to settle to the ground state constrained by screened Coulombic repulsion \cite{haider2009controlled,huff2018binary,rashidi2018initiating} lends this technology well for \gls{fcn}, a class of devices that employ electric \cite{hennessy2001clocking,lent2003molecular} or magnetic field effects \cite{bernstein2005magnetic,alam2010experimental} for operation.
\gls{sidb} logic components demonstrated experimentally by Huff \etal{} \cite{huff2018binary} provide an example of this by taking advantage of the charge sharing behavior between closely-situated \glspl{sidb} to represent binary logic states, wherein the position of a charge in pairs of \glspl{sidb} can be used to encode bit information. A wire can be constructed by placing these \gls{sidb} pairs in series. The study also experimentally verified an OR gate no larger than \SI{5}{\nm} by \SI{6}{\nm} with two \gls{sidb} pairs at the top as logic input and one \gls{sidb} pair at the bottom as logic output, demonstrating logic gates at the atomic scale.

Rapid exploration of \glspl{sidb} outside of experimental laboratories was enabled by SiQAD \cite{ng2020siqad}, a \gls{cad} tool which offers design features for \gls{sidb} layouts and calibrated simulation models. Multiple \gls{bdl} logic gates and circuits have been proposed based on SiQAD verification \cite{ng2020siqad, chiu2020poissolver, ng2020thes, vieira2021novel, bahar2020design}. The authors have experience designing such layouts via SiQAD \cite{ng2020siqad, ng2020thes} and find this to be a time consuming process which requires a significant amount of trial and error.
Designing such layouts via SiQAD is an inherently time-consuming process that requires an understanding of the underlying physics of the technology to be able to navigate the complex, iterative process \cite{ng2020siqad, ng2020thes}.

Electronic Design Automation (EDA) software was essential for the scaling and robust conceptualization of modern semiconductor chip designs. \cite{Scheffer2006} Currently, there exists no automated \gls{sidb} gate and circuit design framework. Other \gls{fcn} technologies generally have well-defined logic unit cells and elementary gates which lower the barrier for logic abstractions and design automation frameworks \cite{trindade2016placement, walter2019fiction, walter2020design}. On the other hand, \glspl{sidb} offer much flexibility with quantum dot placement; this enables the creation of compact and expressive \gls{sidb} gates at the expense of physical complexity and an increased initial learning curve for the designers. This can be managed to an extent with physically-informed design rules, but this does not fully offset the need for part tweaking necessitated by surface and environmental variations in practice. Automated design tools are necessary for the quick prototyping that is required for \gls{sidb} technology to scale to a practically large and useful scope.

The robust design of an SiDB layout that implements the logic of a truth table with an arbitrary number of inputs and outputs is an inherently challenging hyperdimensional optimization problem. Several attributes of the SiDB framework contribute to the difficulty of conceptualizing and implementing a robust optimization algorithm that functions across scales. Firstly, the problem's primary objective function is a discrete logical truth table, meaning that no gradients are available to help guide the parameters towards the optimal set for arbitrary inputs. Furthermore, the objective function is capable of providing useful feedback only at the very end of the optimization process, as there is no deterministic way of determining if a certain track of SiDB placement decisions will lead to a successful design. Binary feedback for the optimization algorithm is available only after the final layout design is determined--either all steps leading up to the final solution were successful in satisfying all rows the truth table, or not. 
Secondly, the optimization space is highly counterintuitive due in part to inter-DB coupling and quantum mechanical phenomena. It also presents many singularities of invalid layouts (due to the existence of non-experimentally characterized positive charge states), as well as sparse singularities of working layouts. \cite{ng2020siqad}
Both of these optimization difficulties are faced by drug molecule discovery optimization, where the relationships between the structures of atoms within and across functional groups are not known until after the final evaluation of the molecule. \cite{Schneider2018, Popova2018} The drug discovery community has found success with reinforcement learning algorithms for traversing optimization spaces with characteristics similar to those of SiDB placement \cite{Popova2018}, indicating that similar techniques are likely to be successful for SiDB layout optimization.

The numerous factors that introduce significant difficulty in the optimization of SiDB layouts single out reinforcement learning as an optimization framework that is uniquely suited for this problem. Double-deep Q-learning \cite{hasselt2016} is chosen as the optimization algorithm for SiDB layout optimization given the framework's proven ability to determine robust policies in domains with partially-observed, low-dimensional state spaces, similar in kind to that of the SiDB design problem. Reinforcement learning (RL) in general is ideally positioned to find optimal solutions in the SiDB design space, where only the final outcome of the optimizer that can offer a reward directly useful for learning optimal layouts \cite{Sutton2018}: whether the final layout of SiDBs produces the desired logic. Since RL is capable of determining the value of actions in the early stages of optimization with respect to achieving the end goal, by learning the expected value of the maximum sum of all future rewards given a particular at any step of optimization, the algorithm can function within the low-affirmation confines of SiDB layout design.
Double-deep Q-learning in particular is well-suited for SiDB design optimization because it helps correct the overestimation of action values---a flaw inherent to many RL algorithms in which the value of actions that lead to positive rewards in the short term are over-estimated.\cite{hasselt2016} This flaw is exacerbated by the rare distribution of rewards throughout SiDB placement optimization, and would result in non-convergence if not addressed by double-deep Q-learning.

In this Article, we introduce the first automated designer for SiDB logic layouts. The robust reinforcement learning framework we built permits design across all scales of layouts demonstrated thus far in the literature, as well as for an arbitrary number of inputs and outputs. It is demonstrated that the designer excels in discovering design strategies for a wide range of logic type without a change of hyperparameters and that the designer demonstrates robustness through convergence to the same design policy regardless of the random seed used in the stochastic processes of the algorithm. This automated designer has demonstrated its utility as a practical tool, as it has already enabled the creation of the Bestagon gate library \cite{walter2022hexagons}, an \gls{sidb} gate library that paves the groundwork for higher level \gls{sidb} design automation exploration.

\section{Results}

To utilize the power of a reinforcement learning agent as an automated design tool, it is necessary to first cast the SiDB layout design process as a reinforcement learning problem.  This entails framing the design space as a Markovian \textit{state space}, $S_t$, at each time step $t$, choosing the action space, $A_t$, update rules, and a method for learning the maximum cumulative value of all future actions given the choice of each possible action at general time step $t$.\cite{Sutton2018} The reinforcement learning agent starts with an empty design area between the input and output SiDBs, and at each step of the Markovian double-deep Q-learning process it places SiDBs in an attempt to either explore the design space or exploit the best design policy learned up until the current time step, $t$.

\begin{figure*}[ht]
  \centering
  \includegraphics[width=1.0\textwidth]{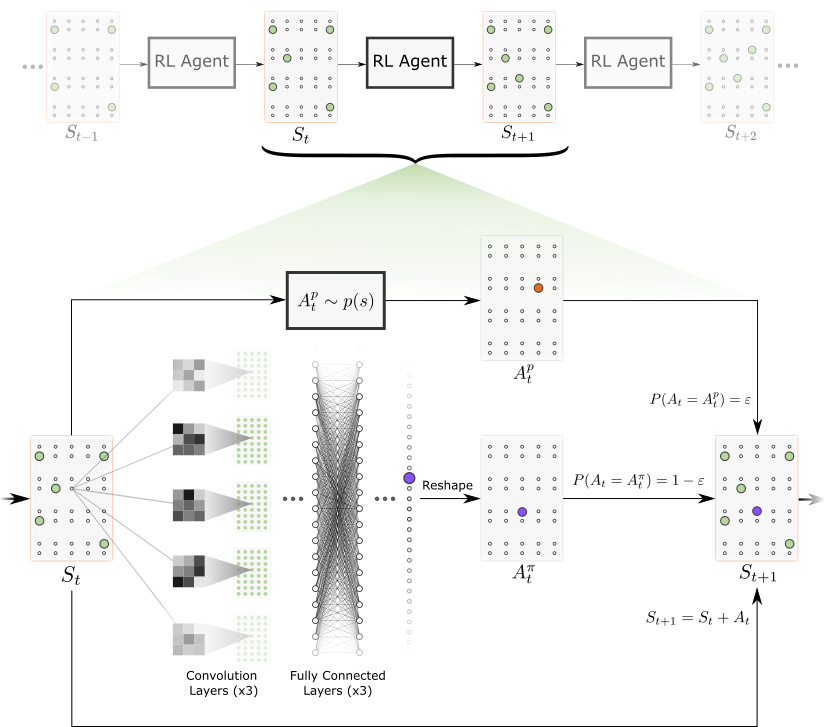}
  \caption{Overview of the double-deep Q-learning iterative SiDB dot placement automated design procedure. Starting from a blank design area located between the input and output wires of the logic circuit, the reinforcement learning agent is iteratively called to place SiDB dots until either a new working layout is discovered or the maximum number of allowable SiDB dots are placed. The reinforcement learning agent places a new SiDB dot based on the recommendation of either a uniformly random exploratory policy, $A^p$, or a neural network-learned greedy exploitative policy, $A^\pi$. Initially, all actions are exploratory to train the neural network, and this is slowly annealed as the agent gains experience. The neural network is composed of three convolutional layers followed by three fully connected layers. It is trained to learn the maximum attainable future rewards for each action based on the inputted state space at time $t$, $S_t$.
  }
  \label{fig:fig1}
\end{figure*}

An iterative SiDB addition scheme was chosen for the design tool, as illustrated in Fig. \ref{fig:fig1}. The action space at any time step $t$ is chosen to be the addition of an SiDB at any valid location within a user-defined area, which is usually defined between the input perturber SiDBs and the output SiDBs, which are oftentimes in proximity to a \textit{weak spring}. \cite{ng2020siqad, ng2020thes} Valid SiDB locations are defined as any point on the lattice within the design area that is not already occupied by an SiDB, which would have been placed in a prior step, or is immediately adjacent an existing SiDB. SiDBs are not allowed to be adjacent to each other because this leads to degenerate states and an over-saturated potential landscape, which has not been experimentally validated to be constructive towards useful SiDB logic layouts. The action to be taken in the presence of action space $S_t$ is either determined by the exploration policy, $A_t^p$, if the agent is exploring the design space, or otherwise by the learned action policy, $A_t^\pi$. Initially, 100\% of actions are exploratory, and this is linearly annealed to the minimum rate, 10\%, by the time that 80\% of design time has elapsed. $A_t^p$ chooses from the valid action space of state $S_t$ through a uniformly random policy. $A_t^\pi$ greedily chooses the action corresponding to the greatest value outputted from a convolutional neural network that is fed $S_t$ as input, which is trained to estimate the maximum attainable cumulative future rewards for each choice in the action space. In this manner, SiDBs are iteratively added to the initially-blank design space, with the process terminating either upon the discovery of a newly-encountered working layout, or after a pre-determined maximum number of SiDBs are placed and no solution is found (which corresponds to a \textit{losing trajectory}).

A neural network is trained to predict the maximum cumulative value of all future possible actions for an arbitrary state $S_t$ by training it based on prior experiences of rewards that were administered to the network, which are accumulated as a result of both the exploratory and exploitative policies. Towards the beginning of the reinforcement learning process, the neural network is seldom exploited, and experiences are accumulated through exploration of the design space. The neural network gradually becomes more accurate, but a 100\% greedy policy is never pursued. A hyperparameter grid search indicates that a 10\% minimimum exploration rate is beneficial to maintaining an output of new, diverse SiDB designs in the latter stages of the reinforcement learning algorithm and preventing overfitting to the same handful of high-performing, already-discovered working layouts. The neural network architecture is chosen using a grid search, with the final design consisting of one input shape-preserving, padded convolutional layer, two non-padded convolutional layers, and three fully-connected layers, where the output is of the same size as the entire state space. The convolutional layers encode spatial relationships amongst the SiDBs, which are then interpreted by the dense fully-connected layers.

\begin{figure*}[ht]

\subfloat[Average reward administered over training epochs for various two-input, one-output logic gates, for both the reinforcement learning algorithm and a uniformly random control policy.]{%
  \includegraphics[clip,width=0.8\textwidth]{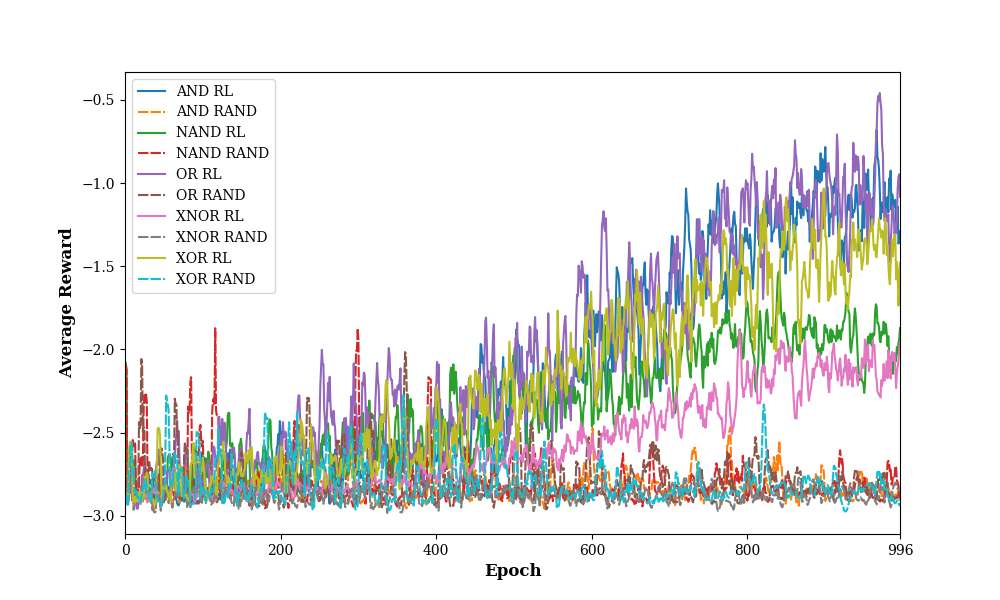}%
  \label{fig:fig2a}
}

\subfloat[Average reward administered over training epochs for a two-input, two-output half-adder, for both the reinforcement learning algorithm and a uniformly random control policy.]{%
  \includegraphics[clip,width=0.8\textwidth]{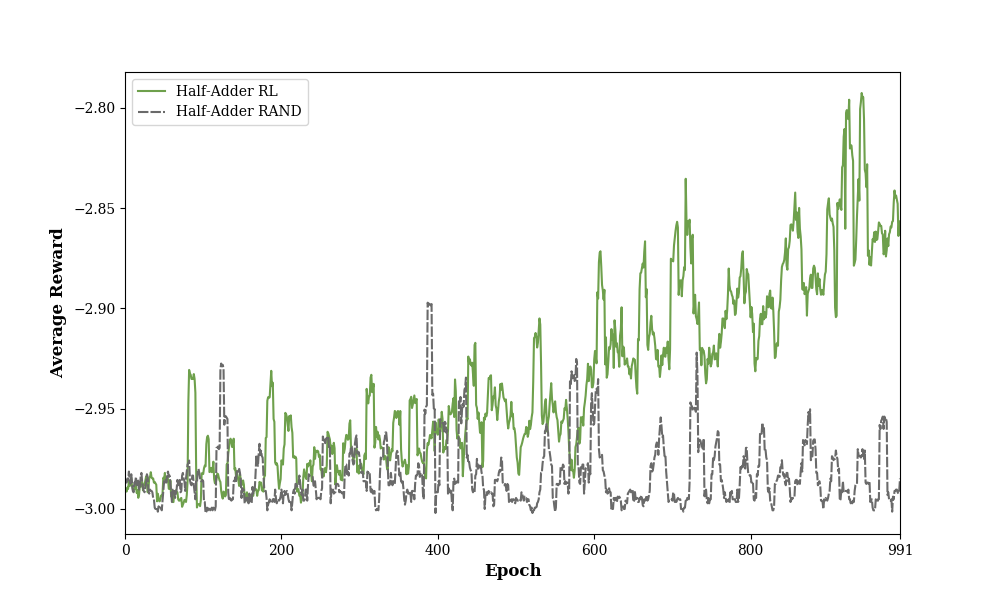}%
  \label{fig:fig2b}
}

\caption{Average rewards gained by the reinforcement learning agent as a result of its actions, along with the results from a uniformly random control policy. The presence of the saturation of administered awards indicates that the convergence to an optimized design policy.}

\end{figure*}

The rewarding strategy is selected such that it is robust across all scales of layout areas and truth table sizes. Rewards are selected to have values between -1 and 1, facilitating training by limiting the scale of backpropagated prediction error gradients \cite{Mnih2015} and allowing for the use of the same learning rate regardless of the number of truth table rows. To encourage the reinforcement learning agent towards completing the entire truth table through each SiDB placement, the addition of a truth table row as a result of an action (i.e., the addition of an SiDB), is rewarded by 0.4 divided by the total number of rows in the truth table. Similarly, to prevent the erasure of progress through the addition of poorly placed SiDBs, a reward of -0.4 divided by the total number of rows in the truth table is adiministered for each truth table row loss. Each placement step, regardless of outcome, is also given a reward of -0.75 divided by the total maximum number of steps in an SiDB placement sequence to encourage efficient choices. Finally, a reward of 1.0 is awarded for newly discovered working layouts, which promptly terminates the placement sequence and begins the next one. If a previously-discovered layout is found, the 1.0 award is not administered and the placement sequence continues, in search for more working layouts.

The reinforcement learning automated layout design algorithm is effective in finding successful design strategies for a diverse range of layouts that span area scales and truth table logic levels of complexity. As illustrated in Fig. \ref{fig:fig2a}, the agent is capable of reliably finding design policies that maximize the average reward administered per epoch for two input, one output logic circuits. In Fig. \ref{fig:fig2b}, it is demonstrated that the same agent, utilizing the same neural network architecture and with unchanged hyperparameter values, is capable of determining a working design policy for the larger and more logically complicated two-input, two-output half adder circuit. The average reward is directly correlated to the number of solutions found by the design tool, with a saturated average reward distribution indicating the convergence to an optimized design policy.

\begin{figure*}[ht]
  \centering
  \includegraphics[width=0.8\textwidth]{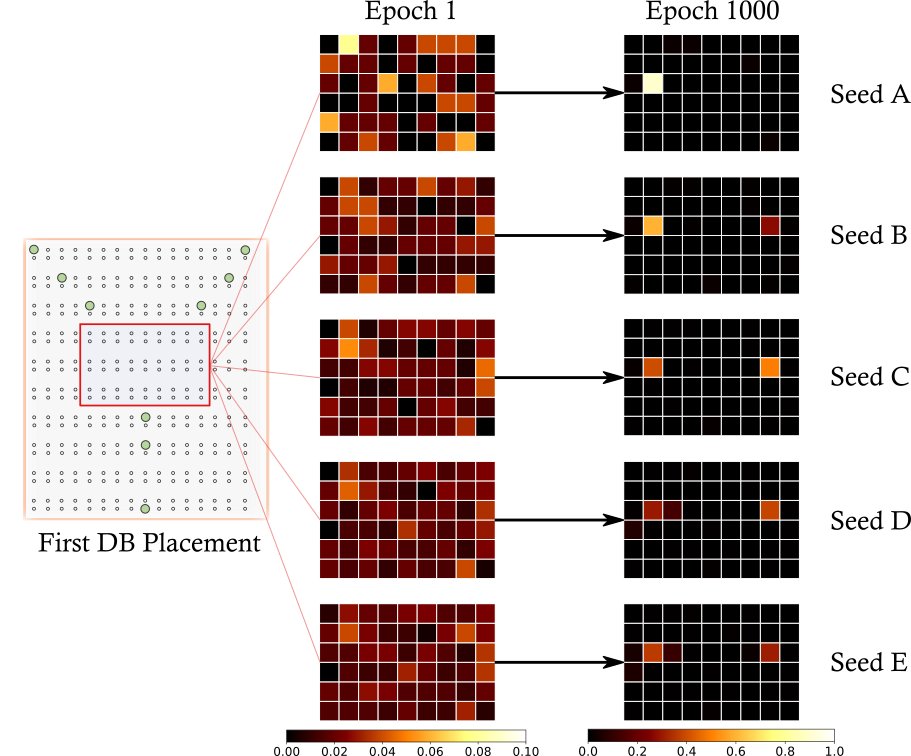}
  \caption{Overview of the evolution of five agents with different random initialization seeds converging to the same optimized design policy. For the two-input, one-output AND gate, the five agents' SiDB dot placement probability distributions at epoch $1$ (before training) are all different and are widely spread out across the entire design canvas. After $1,000$ epochs of designing and training, four of the five agents converge to the same first-dot placement probablility distributions, and the fifth converges to the same policy through a different distribution. Namely, the automated designer learns to place two SiDB dots symmetrically around the center axis, towards the input wires---a design choice that is very familiar to the human-designed circuits presented in the literature. \cite{ng2020siqad}
  }
  \label{fig:fig3}
\end{figure*}

Regardless of the seeds used to initialize the stochastic processes involved in training the reinforcement learning agent, the algorithm has demonstrated a propensity to converge to the same set of optimal design solutions. As illustrated in Fig. \ref{fig:fig3}, given five different sets of random initialization seeds for the neural network optimization initialization and the reinforcement learning agent's exploration policy, it is demonstrated that different starting policies will eventually converge to the same design methodology. This indicates that the design tool is robustly capable of converging to an optimized action policy, rather than haphazardly finding a set of actions that happen to work. Furthermore, the first SiDB placement policy that was determined by all five agents in Fig. \ref{fig:fig3} are very similar to those determined by human designers in previous works, namely placing two SiDBs symmetrically around the center axis of the design canvas, towards the input wires.

\section{Discussion}

The \gls{sidb} platform technology offers unparalleled flexibility in logic gate design in comparison to other \gls{fcn} technologies; however, harvesting this flexibility requires a great deal of experience and trial and error on the researchers' part. Considering that the silicon surfaces in practical real world usage may have various defects\cite{huff2019electrostatic} that circuits may have to account for, it is evident that it is unscalable to rely solely on manually designing \gls{sidb} logic layouts.
From the authors' \gls{sidb} logic design experience\cite{ng2020siqad, ng2020thes}, the generated logic gates are often more compact than their hand-designed counterparts with an order of magnitude time savings. Further, when considering large scale \gls{sidb} systems, the automated designer can also account for physical parameter variations due to differing dopant concentrations, physical defects, or other external influences, something that manual design works have yet to develop design rules or procedures for.
As previously mentioned, the automated designer has already seen practical utility in assisting with the creation of the Bestagon gate library \cite{walter2022hexagons}.
This work paves the way for the proliferation of \gls{sidb} platform technology by introducing a fast and robust automated design tool to experienced and inexperienced quantum circuit designers alike---a crucial piece required to elevate the technology from a research curiosity to an industry staple.


\section{Data availability}

All code and data will be made available at https://github.com/rclupoiu/db-layout-designer upon publication of this manuscript.

\bibliographystyle{naturemag}
\bibliography{refs}

\section{Acknowledgements}
We acknowledge the support of the Natural Sciences and Engineering Research Council of Canada (NSERC).

Robert Lupoiu is supported by a graduate fellowship award from Knight-Hennessy Scholars at Stanford University.

\section{Competing interests}

The authors declare no competing interests.

\end{document}